\documentclass[onecolumn]{elsart3p}

\usepackage{amssymb,subfigure}
\usepackage[latin1]{inputenc}
\usepackage[frenchb,english]{babel}
\usepackage{times,amsmath,amsfonts,graphicx,scalefnt,fancyvrb}
\usepackage{color,listings} 
\def\xmllab{XMLlab}

\begin{document}

\begin{frontmatter}
\title{\xmllab\   \thanksref{label1}: multimedia publication of simulations applets\\using XML and Scilab}
\thanks[label1]{http://xmllab.org}

\author{Stéphane Mottelet}
\address{Laboratoire de Math\'ematiques Appliqu\'ees de Compi\`egne, 
D\'epartement de G\'enie Informatique, 
Universit\'e de Technologie de Compi\`egne, 
BP 20529, 60205 COMPIEGNE CEDEX, FRANCE}
\author{André Pauss}
\address{UMR Génie des Procédés Industriels, Département de Génie Chimique, Universit\'e de Technologie de Compi\`egne, BP 20529, 60205 COMPIEGNE CEDEX, FRANCE}
\ead{stephane.mottelet@utc.fr}


\author{}

\address{}

\begin{abstract}
We present an XML-based simulation authoring environment. The proposed description language allows to describe mathematical objects such as systems of ordinary differential equations, partial differential equations in two dimensions, or simple curves and surfaces. It also allows to describe the parameters on which these objects depend. This language is independent of the target software and allows to ensure the 
perennity of author's work, as well as collaborative work and content reuse. The actual implementation of XMLlab allows to run the generated  simulations within the open source mathematical software Scilab, 
either locally when Scilab is installed on the client machines, or on thin clients running a simple web 
browser, when XMLlab and Scilab are installed on a distant server running a standard HTTP server. 
\end{abstract}

\begin{keyword}
simulation markup language\sep interoperability \sep multimedia publication
\PACS 
\end{keyword}
\end{frontmatter}

\section{Introduction}
\label{}

The need to use a simulation tool is in most cases an answer to simple statements : the user has some equations modeling a physical system. He wants to solve them, and if possible to be able to easily change some parameters to see how they influence the results of the simulation and finally save the parameters and the results (e.g. in a format readable by a spreadsheet application). 

The educational benefit of using simulations, when an adequate tool is used, is not to be discussed here. But there are very different steps in the development
of a simulation. Once the equations are stated, you firstly have to make them fit to a particular software, provided this software is adequate to the
disciplinary field of the phenomenon. This first step is not time-consuming compared to the time which is always spent to develop a graphical user interface.
The author will spend the greater part of his time to polish the interface, although he could have spent this time to work on another simulation. Moreover, the more the applet will be polished to fit a particular case, the less it will be reusable in another close context.  
The World Wide Web is a place where a lot of good quality JAVA applets can be found, but these applets are always difficult to reuse in the context of a 
particular course, because modifying them (when the author makes the source code available) needs abilities in a low level language (JAVA, C++, C), or a 
high level script language such as the one used by Matlab or Scilab (\cite{scilab,scilabengl,scilabopen}).

This kind of work is the concern of craftsmen, and not of an industrial approach. The author's work is not reusable in general and its perennity is not
guaranteed, because the work relies on an application using a proprietary format, and last but not least, the work is exchangeable exclusively with authors
using the same application (and sometimes the same version).

People working on modern documentary applications have already made this analysis, and this can be seen with the exponential growth of the number of
applications of the XML markup language (\cite{xml}). In the field of simulation applets this work has just 
begun. One can cite, in the field of biology and chemistry, the work of a consortium of academic people and authors of 
simulation software which has lead to \verb1sbml1, and exchange markup language modeling biological and chemical 
systems (with kinetics) using XML (\cite{sbml,sbml2}).

Another project is \verb1xmds1, a tool also based on XML allowing to generate Scilab, Matlab or C++ code (but without any graphical user interface)   allowing to simulate deterministic or stochastic systems (\cite{xmds}). Compared to our approach, which will detailed in this paper, the weakness of this tool is a lack of structuring in the description language (the level of structuring is not deep enough compared to what XML allows to do).  

Concerning other serious XML based simulation modeling projects, we can refer to \cite{bgml}, where the use of XML markup
to describe bond graph models is considered. This paper has an excellent introduction recalling the 
essential features of XML technologies relevant for the description and the processing of bond graph models, which
is also relevant for others high level approaches for the description of systems, like ours. 

In the scope of the \xmllab\ project, we have chosen to show the benefits of an approach where the content and the form are well differentiated and are the concern of different people:
\subsection*{The content} The content of simulation resides in the equations of the phenomenon, their description, the associated parameters and their thematic organization. The description of the content is the concern of the author. 
\subsection*{The form} The form resides in the graphical user interface, the various ``widgets'' and menus which reinforce the user-friendliness of the final applet, and the visualization tools (static curves, animations, sounds). This part of the applet code is the concern of a high level developer, or the concern of a tool able to generate this code automatically from the description of the content. This is the option we have chosen.  

\bigskip The choice of adequate numerical methods is another concern. The author is not necessarily able to make this choice himself, that's why this choice has to be done automatically, knowing which method is the most adequate to solve a given type of equation.

The purpose of the XMLlab project was not to define a description language by its own, hence we have chosen a particular ``target'' application in the early phases of the project. This application is Scilab, an {\em open source} software developed since 1990 by researchers of INRIA and ENPC. Our goal was to develop a complete ``compilation chain'', allowing to transform the source XML documents into executable scripts interpreted by the target application. Moreover, the choice of Scilab is motivated by the fact that this software allows to use the Tcl/Tk script
language (\cite{tcltk,tcltkbook}) to generate graphical user interfaces. We also use the Tcl/Tk message passing system
in the XMLlab WebServer, which allows to dynamically publish the simulations on the web. In fact this paper extends \cite{xmllab2004}, since the new 
features of \xmllab\ generalize its multi-medium nature.

\section{The structure of an \xmllab\ simulation}

As specified in the \xmllab\ DTD, a simulation can be divided in a certain number of conceptual elements: parameters, mathematical models of 
objects (time-dependent or not) and finally a display element to output the results of the simulation.
\subsection{Parameters}
They are the parameters of the phenomenon and of the mathematical model. The goal is to allow the user of the simulation to make them vary by 
means of the interface which will be generated by the compilation chain. It has to be possible to simply specify if the value of the parameter 
is seen in  the interface, but non modifiable by the user. There must also exist hidden parameters, for internal use only. The parameters are 
grouped into sections, e.g. for an ordinary differential equation, the user may want to 
differentiate the physical parameters of the phenomenon from the resolution parameters (final time, number of discretization steps, etc.). This 
logical structuring can be then used to graphically structure the interface.

\medskip
\begin{itemize}
\medskip\item{\bf Scalars and matrices} : with \xmllab\ , a simulation the parameters can be scalars or matrices. The minimum and maximum value of a scalar can be given, the type of ``widget'' to use (a slider, or a simple entry field where the user can modify the value). Each parameter must have symbolic name which can be reused in the description of the mathematical model, and a default value, which will be used for the first run of the simulation.
\medskip\item{\bf Databases} : the user can store many instances of a parameter group. This allows, e.g. in chemistry, to build a small database of different acids and alkali, by storing their parameters (acidity constants, charge, etc.). The database can then be used to generate a menu allowing  to  choose a given parameter group in the interface.
\end{itemize}
\subsection{Objects with a mathematical model}
We now deal with the equations of the phenomenon to be simulated. There are elements of different levels. 
\begin{itemize}
\medskip\item{\bf Domains} : the most simple describe intervals of $\mathbb{R}$ or domains of $\mathbb{R}^2$. They are simple closed intervals of the type $[a,b]$ (where the bounds may depend on parameters described in the previous section) or two dimensional domains. The latter can be rectangles defined by a Cartesian product of two intervals, or general domains defined by the form of their boundary by parametric curves (we will discuss curves in the next item). The user can  precise the way these domains have to be discretized, if applicable  (number of discretization points, linearly or logarithmically).
\medskip\item{\bf Curves and surfaces} : non-parametric curves can be described like this,
$$
y=f(x), \quad x\in[a,b].
$$
This definition reuses an interval. This way, it is possible to define many curves referring to the same interval. Parametric curves can be also defined like this,
$$
x=f(t),\;y=g(t),\;t\in[a,b].
$$
The surfaces can also be of parametric or non-parametric type, namely
$$
z=f(x,y), \quad (x,y)\in\mathcal{D},
$$
where $\mathcal{D}$ is a domain of $\mathbb{R}^2$, or
$$
x=f(u,v),\,y=g(u,v),\,z=h(u,v),\;(u,v)\in \mathcal{D}.
$$
The parameters defined in the previous section can be used at any level, in the definition of domains or in the equations themselves.
\medskip \item{\bf Ordinary differential equations} : one can describe systems of ordinary differential equations, e.g.
$$
\frac{d}{dt}x(t)=f(x,y,t),\;
\frac{d}{dt}y(t)=(x,y,t),\;
t\in[a,b],
$$
with given initial conditions $x(a)=x_a$ and $y(a)=y_a$. \xmllab\ allows to keep the natural description, without having to reformulate 
each unknown $x(t)$ or $y(t)$ as the element of a vector $X(t)$ with $x(t)=X_1(t)$ and $y(t)=X_2(t)$. The chosen description model 
consists in:
\begin{itemize}
\item A time interval (here $[a,b]$)
\item A list of states. For each state (here $x$ or $y$), its time-derivative and its initial value are given.
\item A list of outputs. They are the observations which can be computed by using the states, e.g. $z(t)=x(t)+y(t)$. 
\end{itemize}
\medskip\item{\bf Non linear equations } : general systems of non-linear equations can be described, namely 
$$f(x,y,z,t,\cdots)=0,\;
h(x,y,z,t,\cdots)=0,\;
\cdots
$$
or curves defined by an implicit equation  of the form
$$
f(x,y)=0, x\in[a,b],
$$
This kind of equation is used in the modeling of acid-alkali titration.
\medskip\item{\bf Partial differential equations } :  \xmllab\ allows to describe partial differential equations of diffusion type, namely
$$
\left\{\begin{array}{c}-\operatorname{div}\left(P\operatorname{grad}u\right)(x)+c(x)u(x)=f(x),\; x\in \Omega,\\
+\text{\em\ boundary conditions}
\end{array}\right.
$$
The domain $\Omega$ is described from its boundary (parametric curves, defined earlier). We will give some details on the numerical methods in the next section. Here again, the parameters can be used at any level, in the definition of the domain $\Omega$ or in the physical data (diffusion matrix $P$, source term $f(x)$, proportional coefficient $c(x)$).
\end{itemize}

\subsection{Results display}
We use a classical hierarchical description, using windows and systems of axes, where the user just has to precise what he wants to display by making 
reference to objects defined in the ``Mathematical models'' section.
\begin{itemize}
\medskip\item{\bf Windows} : A window contains systems of axes. The user just has to specify how they have to be placed if they are more than one (the window is divided within its height and width).
\medskip\item{\bf System of axes } : the user has to specify if the system is two or three dimensional. Each system of axes contains some references to what has to be represented.
\medskip\item{\bf Objects to be represented graphically} : reference can be made to a curve, to a surface or to the state of an equation,  by means of its symbolic name. The chosen structure  allows to greatly simplify the number of different elements. For example, a surface can be referenced in a two or three dimensional system of axes. In a three dimensional system a perspective projection is used, although in two dimensions we use a pseudo-color planar representation. In both cases, the reference to the surface is made identically, only the the ``parent'' context is changing.
\end{itemize}
\medskip
\begin{figure}
\begin{lstlisting}
<?xml version="1.0" ?>          
<!DOCTYPE simulation PUBLIC "-//UTC//DTD XMLlab V1.6//EN" 
          "http://www.xmllab.org/dtd/1.6/fr/simulation.dtd">
<simulation>
  <header>
     ...
  </header>
  <notes>
     ...
   </notes>
  <parameters>
     ...
  </parameters>
  <compute>
     ...
  </compute>
  <display>
     ...
  </display>
</simulation>
\end{lstlisting}
\caption{The outline of a simulation showing the high-level elements of a simulation}
\label{figskel}
\end{figure}
\begin{figure}
\centering\includegraphics[width=2.75cm]{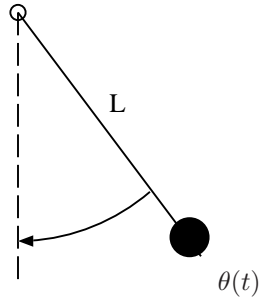} $\theta(t)$
\caption{The pendulum.}
\label{pendulum}
\end{figure}

\section{A typical example of simulation}

We give on the figure \ref{figskel} the skeleton of a simulation document. We will now explain with details how to build this document to describe a small simulation. 

We consider the pendulum depicted on figure \ref{pendulum}. We make the hypothesis that the line connecting the sphere of mass $M$ to the rotation 
axis is of negligible mass compared to $M$. We measure the deviation of the pendulum from the stable vertical equilibrium position by the angle 
$\theta(t)$ positively measured as indicated on figure \ref{pendulum}. If one applies the relations of dynamics for bodies under rotations, we obtain the 
following ordinary differential equation
$$\left\{\begin{array}{rcl}
\ddot\theta(t)  &=& - \displaystyle\frac{g}{L}\sin\theta(t),\quad t\in[0,T]\\
\theta(0)&=&\theta_0,\\
\dot\theta(0)&=&0.
\end{array}\right.
\label{eqPendule:nonlin}
$$
The value of $\theta_0$ gives the initial angular deviation of the pendulum, and we consider that the initial angular velocity is zero. If $\theta_0$ is small, $\theta(t)$ can be 
approximated by $
\phi(t) = \theta_0\cos\left(\sqrt{\frac{g}{L}}t\right)$. 
We want do describe a simulation applet allowing to compare $\phi(t)$ and $\theta(t)$ when $\theta_0$ changes. The graphical output should
plot $\phi(t)$ and $\theta(t)$ for $t\in[0,T]$, and the user should have the possibility to change $\theta_0$ in the interval $[-3.14,3.14]$ by moving the point $(0,\theta_0)$ 
interactively on the curve (the point will be represented with a cross). The XML code fragment containing the description of the parameters is given in
figure \ref{param:pendule}.

\begin{figure}
\begin{lstlisting}
<parameters>
    <section>
        <title>Parameters of the pendulum</title>
        <scalar label="L" unit="m">
            <name>Length of the pendulum</name>
            <value>1</value>
        </scalar>
        <scalar label="g0" unit="ms^-2">
            <name>Gravity</name>
            <value>9.81</value>
        </scalar>
        <point label="point0">
            <x1 label="zero">
                <value>0</value>
            </x1>
            <x2 label="theta_0">
                <value>2</value>
            </x2>
            <constraints>
                <curve ref="segment"/>
            </constraints>
        </point>
    </section>
    <section>
        <title>Resolution parameters</title>
        <scalar label="tf" unit="s" min="0" max="10" increment="1"> 
            <name>Final time</name>
            <value>2</value>
        </scalar>
    </section>  
</parameters>
\end{lstlisting}
\caption{Code fragment containing the description of the parameters of the pendulum simulation}
\label{param:pendule}
\end{figure}
Each pair of \verb1section1 elements allows to group parameters. The \verb1scalar1 element represents a scalar parameter, containing its full name in 
the \verb1name1 element and its initial value in the element \verb1value1. The  \verb1constraint1 element refers to a curve with 
label \verb1segment1, which will be described below. For the final simulation time, corresponding to the parameter with label \verb1tf1 (lines 26 to 29 in figure
\ref{param:pendule}), there
 are bounds (minimum and maximum value) and also a relevant increment. During the compilation phase, the presence of these three attributes values will be
 taken into account and will allow to choose a particular widget in the graphical user interface. 

\subsection{Mathematical models, {\tt compute} element}
Here are the fragments corresponding to the description of  the  $[0,T]$ interval :
\begin{lstlisting}
<defdomain1d label="t" unit="s">
  <name>time</name>
  <interval>
    <initialvalue>0</initialvalue>
    <finalvalue>tf</finalvalue>
  </interval>
</defdomain1d>
\end{lstlisting}
and the XML code fragment describing the differential equation is given in figure \ref{fig:ode}.
\begin{figure}
\begin{lstlisting}
<ode label="pendulum">
  <refdomain1d ref="t"/>
  <states>
    <state label="theta" unit="rad">
      <name>Real solution</name>
      <derivative>theta_dot</derivative>
      <initialcond>theta_0</initialcond>
    </state>
    <state label="theta_dot" unit="rad/s">
      <name>Derivative of the angle</name>
      <derivative>-g0/L*sin(theta)</derivative>
      <initialcond>0</initialcond>
    </state>
  </states>
  <outputs>
    <output label="theta_lin">
      <name>Harmonic solution</name>
      <value>theta_0*cos(sqrt(g0/L)*t)</value>
    </output>
  </outputs>
</ode>
\end{lstlisting}
\caption{Code fragment containing the description of the system of two differential equations of the pendulum.}
\label{fig:ode}
\end{figure}
The \verb1ode1 element (ordinary differential equation) contains an empty element \verb2refdomain1d2 referring to the $[0,T]$ interval 
(defined earlier by the \verb2defdomain1d2 element, referred by  the attribute \verb1ref1), and thus defining the symbolic name of the integration variable, 
a \verb1states1 element containing the description of each state ($\theta$ and $\dot \theta$) in a \verb1state1 element. Each \verb1state1 element contains the 
name of the state, its time-derivative \verb1derivative1 and its initial value \verb1initialcond1. Until now, the content of the \verb1derivative1 element 
is not parsed, and will copied verbatim during the compilation. Since we do not have a dedicated editor, this is easier for the user to type mathematical
formulas like this, but we plan to use Content MathML (see e.g. \cite{mathml}) in the future, which will allow an easier a priori validation of formulas.

The last 
element \verb1outputs1 in \verb1ode1 is a list of outputs, which can be functions of states and/or time. In our example the output explicitly 
depends on time but does not depend on the states. Each state or output has a mandatory attribute \verb1label1 which will be referred in the display section.
\subsection{Graphs, curves and surfaces, {\tt graphs} element}
The code fragment describing the curve mentioned above is given in figure \ref{polyline}. We only need a simple
segment connecting the points $(0,-3.14)$ and $(0,3.14)$. A point lying on this curve
will have its ordinate in the interval $[-3.14,3.14]$. This curve will not be drawn as
it only serves as a constraint for the value of $\theta_0$.
\begin{figure}
\begin{lstlisting}
<graphs>
  <polyline label="segment">
    <vertex x1="0" x2="-3.14"/>
    <vertex x1="0" x2="3.14"/>
  </polyline>
</graphs>
\end{lstlisting}
\caption{Code fragment describing the segment joining the two points $(0,-3.14)$ and $(0,3.14)$.}
\label{polyline}
\end{figure}
It is not needed to write further XML code to construct the curves of $\theta(t)$ and $\phi(t)$ versus $t$, as it is possible
to refer directly to the labels \verb1theta1 and \verb1theta_lin1 in the \verb1<drawcurve2d>1 element (see next section directly below).

\subsection{Display of results, {\tt display} element}
We want to superimpose two curves in the same axes system, and display a movable cross at the $(0,\theta_0)$
coordinate. The XML code corresponding to this is given in figure \ref{displayxml}.
\begin{figure}
\begin{lstlisting}
<display>
  <window>
    <title>Comparison of the two solutions</title>
    <axis2d>
      <drawcurve2d ref="theta"/>
      <drawcurve2d ref="thetalin"/>
      <drawpoints ref="point0"/>
    </axis2d>
  </window>
<display>
\end{lstlisting}
\caption{Code fragment describing the display of the two solutions}
\label{displayxml}
\end{figure}
The \verb1display1 element contains only one \verb1window1 element, containing itself a two dimensional system of axes. The 
two \verb1drawcurve2d1 elements within the same \verb1axis2d1 mean that the curves of $\theta$ and its harmonic version will be superimposed. The
\verb1drawpoints1 element refers to the \verb1point1 element defined before in the \verb1parameters1 element.

\subsection{Remarks}
\begin{figure}
\centering \includegraphics[width=0.5\columnwidth]{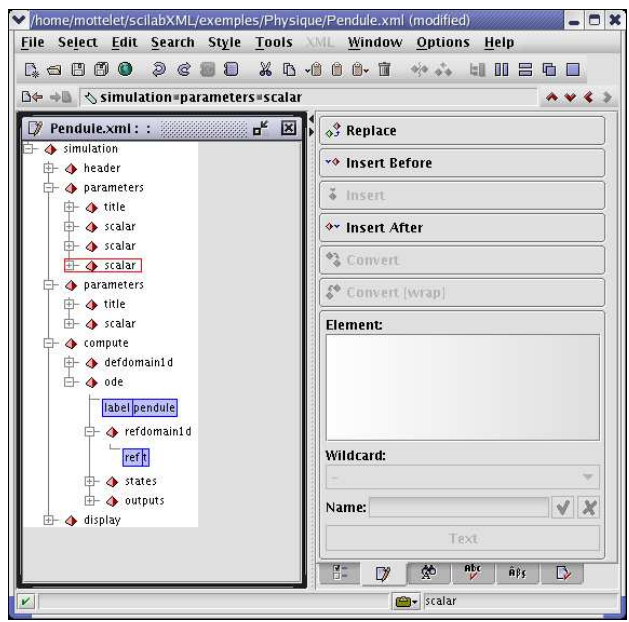}
\caption{\label{xxe}The XML file describing the simulation of the pendulum, seen in the XXE editor, developed by PIXWARE, {\tt http://www.xmlmind.com/xmleditor}}
\end{figure}

The different structuration possibilities are constrained by a DTD (Document Type Definition), allowing an {\em a posteriori} validation of
a simulation, or can be used to constrain the edition of a simulation by means of an XML editor. The figure \ref{xxe} shows the view 
that the user can have of its XML file. 

Within all of the above mentioned elements, some have a particular status. The \verb1name1 and \verb1title1 elements can appear several times, with a different \verb1lang1 attribute (\verb1french1 or \verb1english1 in \xmllab\ 1.3). The goal is to be able to generate from the same XML file two different versions of the ``compiled'' applet, the language to use being specified as a compilation option. 

The \verb1header1 element contains some meta-data such as the name of the author and some keywords. The \verb1notes1 element can appear several times with a different \verb1lang1 attribute and allows to write a few paragraphs of text allowing to describe the simulation and/or to give some help to the user.

\section{The compilation chain}

The compilation chain is entirely based on XML technologies: we use XSL transformations specified in XSL stylesheets (eXtensible Stylesheet Language). These transformations are applied to the simulation file by an ``XSL processor''.
The XSL technology is well known to allow the display of dynamic HTML on the World Wide Web, but it is also well fitted to the automatic generation of scripts. We are here particularly interested in the script language of Scilab or Matlab, and the script language Tcl/Tk, allowing to describe graphical user interfaces.

\begin{figure}
\centering \includegraphics[width=0.5\textwidth]{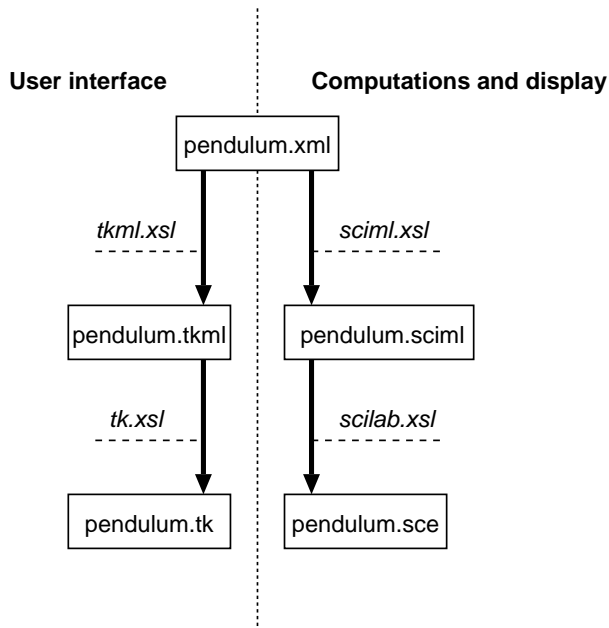}
\caption{Diagram of the compilation chain. The arrows represent the XSL transformations, and the italic names the associated stylesheets.\label{compilation}}
\label{compilationEngl}
\end{figure}

The different phases of the compilation are outlined on the diagram depicted on figure \ref{compilation}. In the bottom of the diagram, the \verb1pendulum.sce1 file is the Scilab script containing all the computation code, and the display of results. The \verb1pendulum.tk1 file contains the Tcl/Tk code of the interface. The diagram illustrates the fact that the transformation is not direct and needs an intermediary step; this particular point needs an explanation. 

To allow an easy maintenance of the XSL stylesheets, and especially to allow a smooth change of target languages 
(Scilab and Tcl/Tk), we have used ``pivot'' XML dialects: the code is generated in a two-step process. We use XSL 
transformation to translate XML input into a pseudo-Tcl/Tk and pseudo-Scilab syntax that are also written using 
XML dialects. Then we use a second pass to serialize the XML pseudo-language into the target language. The advantage 
here is that the second pass captures all the complexities of formatting clean code (syntax) while the first pass 
concentrates on the logical aspects of the translation (semantics). We use the following dialects :

\medskip\begin{itemize}
\item{\bf TKML dialect} : as far as the interface Tcl/Tk code is concerned (left side of the diagram), the intermediary file \verb1pendulum.tkml1 is an XML 
file containing a logical description of the interface. This file contains the description of the different widgets (buttons, etc.) and their placement with respect 
to each other. Then, this intermediate file is finally translated in Tcl/Tk by means of a last XSL transformation.We show on figure \ref{tkml} a small part of the 
generated markup. Almost all the parameters will be appear 
in the graphical user interface as classical entry widgets (corresponding to the ``entry'' widget in Tcl/Tk), where the user has to type the value with the keyboard keys. 
The \verb1<scale>1 element (lines 7 to 10 in figure \ref{tkml}) describes the widget which will be used for the final time of the simulation. Since the original markup describing this parameter in figure
\ref{param:pendule} gives bounds and an increment, the logical way of taking into account these constraints is to use a widget with a moving slider (in the final
transformation we use the ``scale'' widget of Tcl/Tk). Some aspects of the TKML markup are very similar to the XForms markup language, which has been chosen 
by the W3C to develop the next generation of forms technology for the world wide web, see e.g. \cite{xforms}.

\begin{figure}
\begin{lstlisting}
  <page name="id66390" text="Resolution parameters" pady="4">
    <frame packside="top" anchor="n" pady="0" padx="0" fill="x" expand="yes">
      <frame packside="left" padx="5" expand="true" fill="x">
        <label anchor="w" expand="true">
          <text>Final time</text>
        </label>
        <scale anchor="e" variable="tf" state="normal" width="8" from="1" to="10" resolution="1">
          <value>2</value>
          <command>runScilab</command>
        </scale>
      </frame>
    </frame>
  </page>
\end{lstlisting}
\label{tkml}
\caption{TKML intermediate markup corresponding to the definition of the Scilab function computing the right-hand side of the system of ordinary differential equations.}
\end{figure}
\item{\bf SCIML dialect } : for the Scilab code generation, we proceed in the same manner: we first generate an intermediary 
file \verb1pendulum.sciml1, written using a pseudo-Scilab markup, and then transform this file to Scilab code with a last transformation. The figure \ref{sciml}
shows a fragment of the actual content of this file.
\begin{figure}
\begin{lstlisting}
  <function-definition name="f_pendulum">
    <inputs>
      <parm>_t</parm>
      <parm>_X</parm>
    </inputs>
    <outputs>
      <parm>lhs</parm>
    </outputs>
    <body>
      <assign>
        <lhs>t</lhs>
        <rhs>_t</rhs>
      </assign>
      <assign>
        <lhs>theta</lhs>
        <rhs>
          <select matrix="_X" row="1:1" col="1"/>
        </rhs>
      </assign>
      <assign>
        <lhs>theta_dot</lhs>
        <rhs>
          <select matrix="_X" row="2:2" col="1"/>
        </rhs>
      </assign>
      <assign>
        <lhs>lhs</lhs>
        <rhs>
          <list sep=";">
            <parm>(theta_dot)</parm>
            <parm>(-g0/L*sin(theta))</parm>
          </list>
        </rhs>
      </assign>
    </body>
  </function-definition>
\end{lstlisting}
\label{sciml}
\caption{SCIML intermediate markup corresponding to the definition of the Scilab function computing the right-hand side of the system of ordinary differential equations.}
\end{figure}
\end{itemize}

We show on figure \ref{scilabcode} a small part of the generated Scilab script \verb1pendulum.sce1 appearing on figure \ref{compilationEngl}, corresponding to the computation of the right-hand side of the first order
differential equation obtained for the simulation of the pendulum, the call to the ode solver of Scilab and finally the display of curves (for sake of
simplicity we have omitted some parts of the code).

\begin{figure}
\begin{lstlisting}
function [lhs]=f_pendulum(_t,_X)
t=_t;
theta=_X(1:1,1);
theta_dot=_X(2:2,1);
lhs=[(theta_dot);(-g0/L*sin(theta))];
endfunction

// Time
t=linspace(0,tf,200)';
// Script code for the pendulum ode
_X0(1:1,1)=theta_0;
_X0(2:2,1)=0;
_X=ode(_X0,0,t,f_pendulum);
theta=_X(1:1,:)';
theta_dot=_X(2:2,:)';
thetalin=theta_0*cos(sqrt(g0/L)*t);

// Display
plot(t,theta,);
hold("on");
plot(t,thetalin,;
hold("off");
\end{lstlisting}
\caption{Some parts of the generated Scilab script for the pendulum simulation.}
\label{scilabcode}
\end{figure}

\medskip Two ways of distribution can be used: the two Tcl/Tk and Scilab files can be later used without using the XML source and the 
compilation chain (thus protecting the author's work). However, it would be more profitable to the community to release the XML source.

The whole compilation chain together with Scilab (except the XML editor), uses only open-source software packages (\verb1xsltproc1 of the Gnome project, Tcl scripts), and works on any platform (Windows, Mac OS X, Unix). 

\section{The different ways of publishing an XMLlab simulation}
In the previous section, we have described the classical way of publishing an XMLlab simulation,  by using a 
``compilation chain'' allowing to transform the original XML file \verb1pendulum.xml1 to a Scilab executable 
file \verb1pendulum.sce1 and a Tk file \verb1pendulum.tk1 which will be run on a local client machine where Scilab is installed. By using
this kind of publication, the interactivity is maximal; for the pendulum example, the user can move the cross
representing the initial condition and see in real time how the two trajectories diverge (see figure \ref{scilab}).

\subsection{Comments on the example}

\subsubsection{Description of the generated graphical user interface}
\noindent We make some comments on the figure \ref{scilab}. 

\medskip\begin{itemize}
\item {\bf User interface} : the window of the interface has a central space with a Notebook-type widget with tabs and a menu bar:
\begin{itemize}
\item The two tabs named \verb1Parameters of the pendulum1 and \verb1Resolution parameters1 correspond to the two parameters groups specified in the XML file. The user 
just has to select a given tab to display the corresponding parameter group.
\item The \verb1Notes1 tab gives some some information on the simulation, extracted from the \verb1header1 element: name of the author, date and eventual notes 
describing the simulation. The \verb1XMLlab1 tab gives some information on the XMLlab project.
\item The \verb1File1 menu contains two interesting items: ``Save a session'' and ``Load a session''. They allow to save the values of parameters in a text file, and to load them later. It allows to resume a working session (otherwise the Scilab script always starts with the initial values of parameters).
\item The \verb1Languages1 menu allows to switch between the languages of the simulation. In fact, by sake of simplicity we didn't show the textual elements
for each language in the pendulum example, but giving them in all desired languages allows to dynamically switch between languages when running
a simulation. For example, when the default language (English here) and French are to be used, a typical code fragment is the one given in figure
\ref{lang}.
\end{itemize}

\begin{figure}
\begin{lstlisting}
    <section>
        <title>Resolution parameters</title>
        <title lang="french">Paramètres de résolution</title>
        <scalar label="T" unit="s"> 
            <name>Final time</name>
            <name lang="french">Temps final</name>
            <value>2</value>
        </scalar>
    </section>  
\end{lstlisting}
\caption{Code fragment showing textual elements in different languages.}
\label{lang}
\end{figure}
\medskip \item{\bf Graphical window} : the legend of the two curves is taken from the \verb1name1 elements within the states \verb1theta1 and the output \verb1theta_lin1. The abscissa label is the name of the time variable \verb1t1, and the ordinate label is the unit (\verb1unit1  attribute of element \verb1<state label="theta"1). 
This window belongs to Scilab, thus the user has access to the usual menus allowing to save (e.g. in EPS format) or print the figure.
\end{itemize}
\medskip
The user has always the possibility to have access to all variables of the simulation from the Scilab command line (parameters and results of the simulation), which remains available during the simulation.

\subsubsection{Remark on performances}
For this particular example (a system of two scalar ordinary differential equations), the computation time is negligible compared to the time elapsed by drawing the curves, and hence the user can see the immediate effect of the initial angle on the synchronization of the curves. For more intensive examples (e.g. resolution of a partial differential equation), the response time can be greater, but the reactivity of the system is always good, even on a lightweight system (1Ghz Pentium). For these reasons Scilab is really appreciated, because the built-in functions are well optimized (linear and nonlinear equations solving, differential equations, sparse matrix algebra, vectorization of elementary functions for arrays, etc.). Moreover, Scilab loads in a negligible time (compared to the important startup time of recent versions of Matlab, because of the use of JAVA for the user interface).

\newlength{\mylength}
\setlength{\mylength}{0.4\textwidth}
\begin{figure}
$$
\begin{array}{cc}\includegraphics[width=0.4\textwidth]{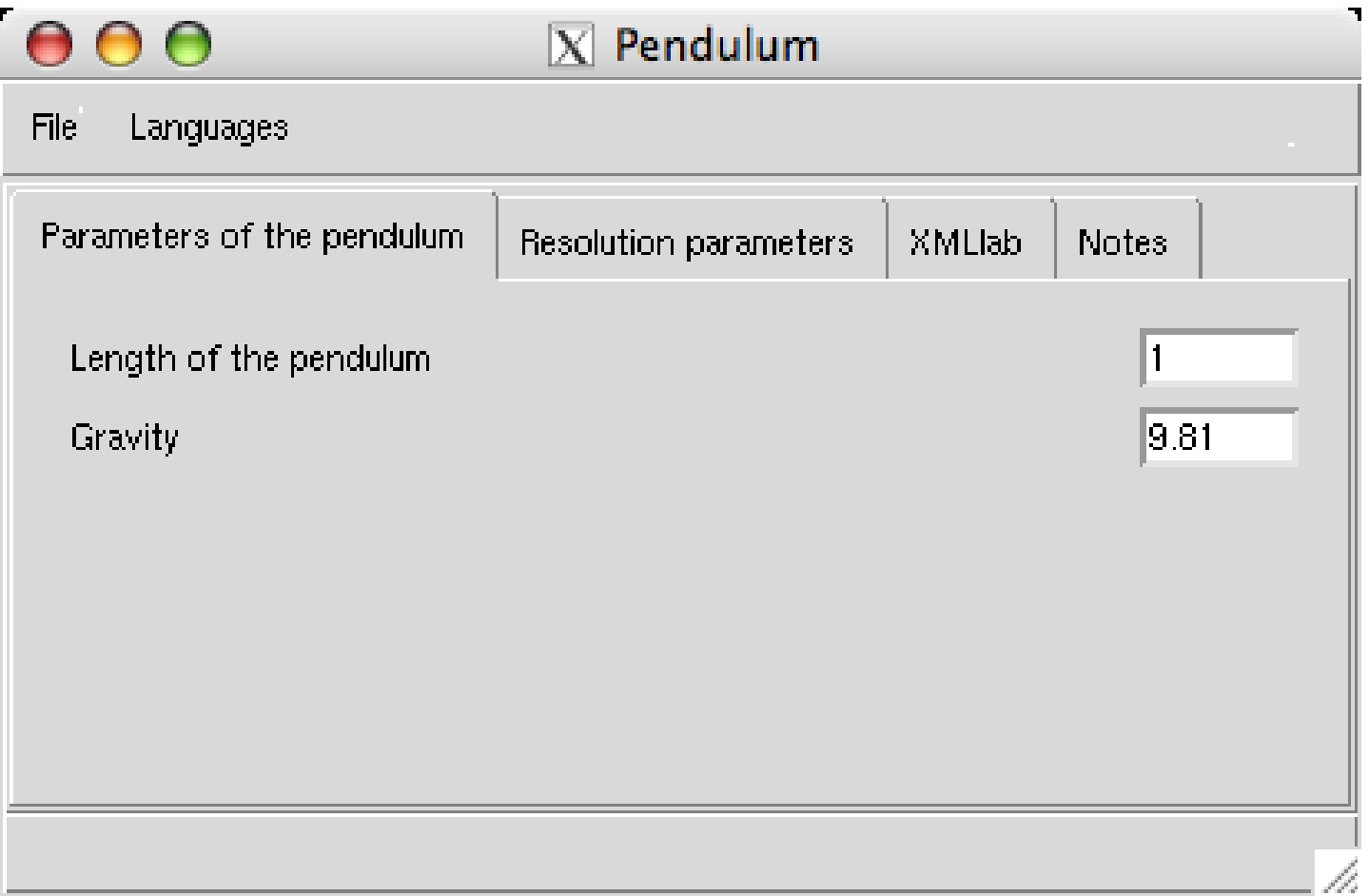} &~~ \includegraphics[width=0.4\textwidth]{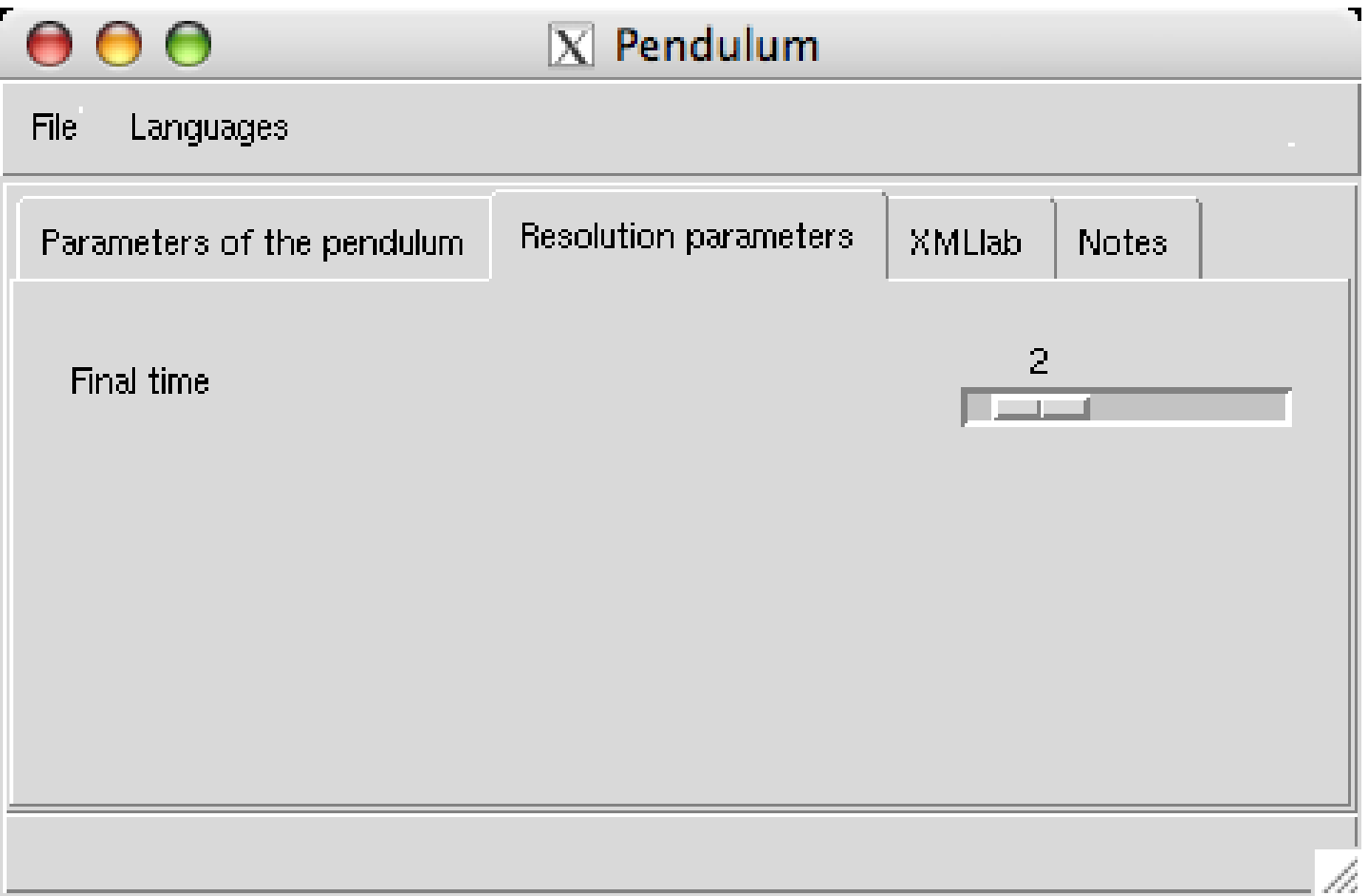}\\
~\\
\begin{minipage}[c]{\mylength}\includegraphics[width=\mylength]{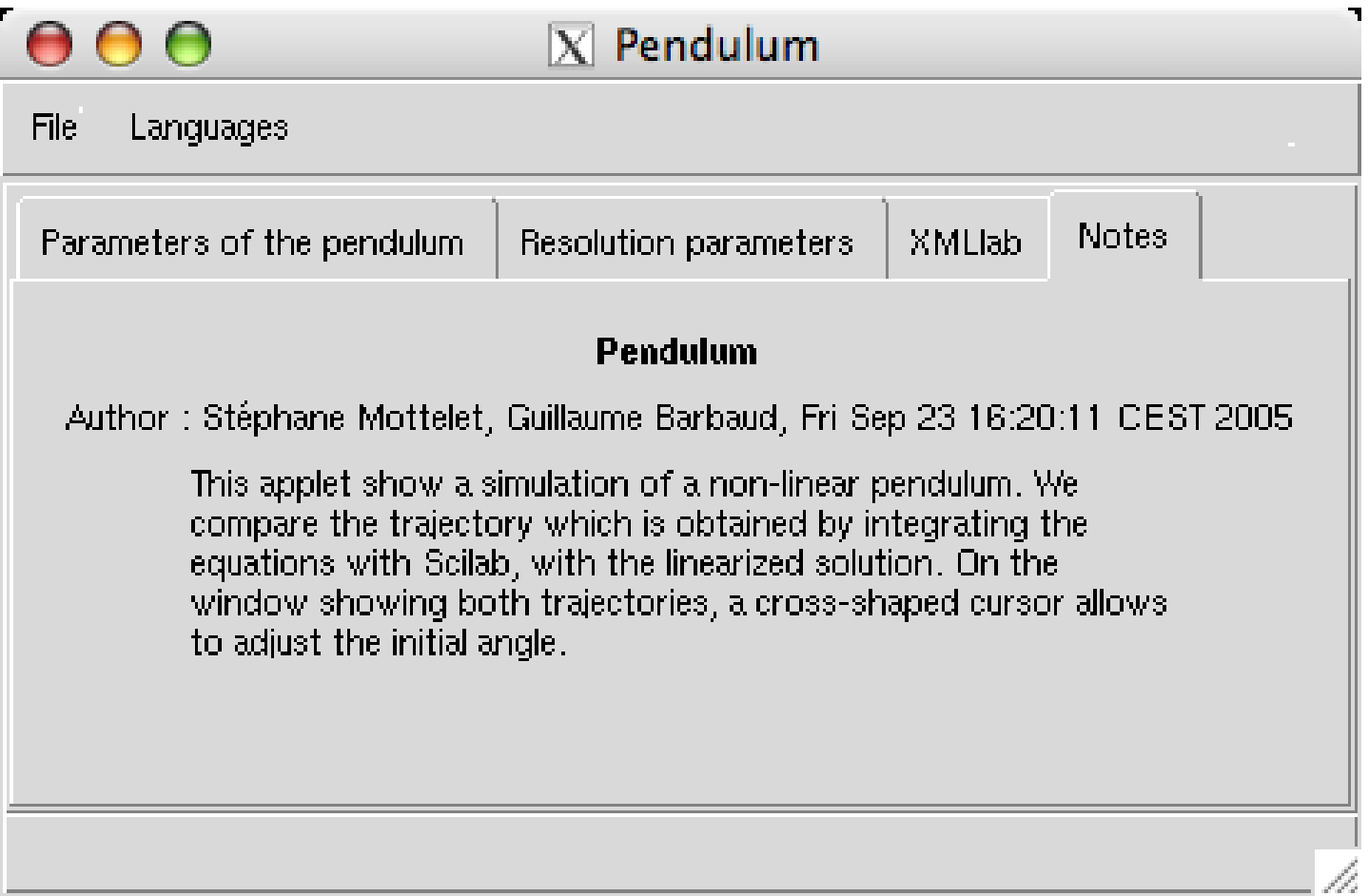}\end{minipage} &~~ \begin{minipage}[c]{\mylength}\includegraphics[width=\mylength]{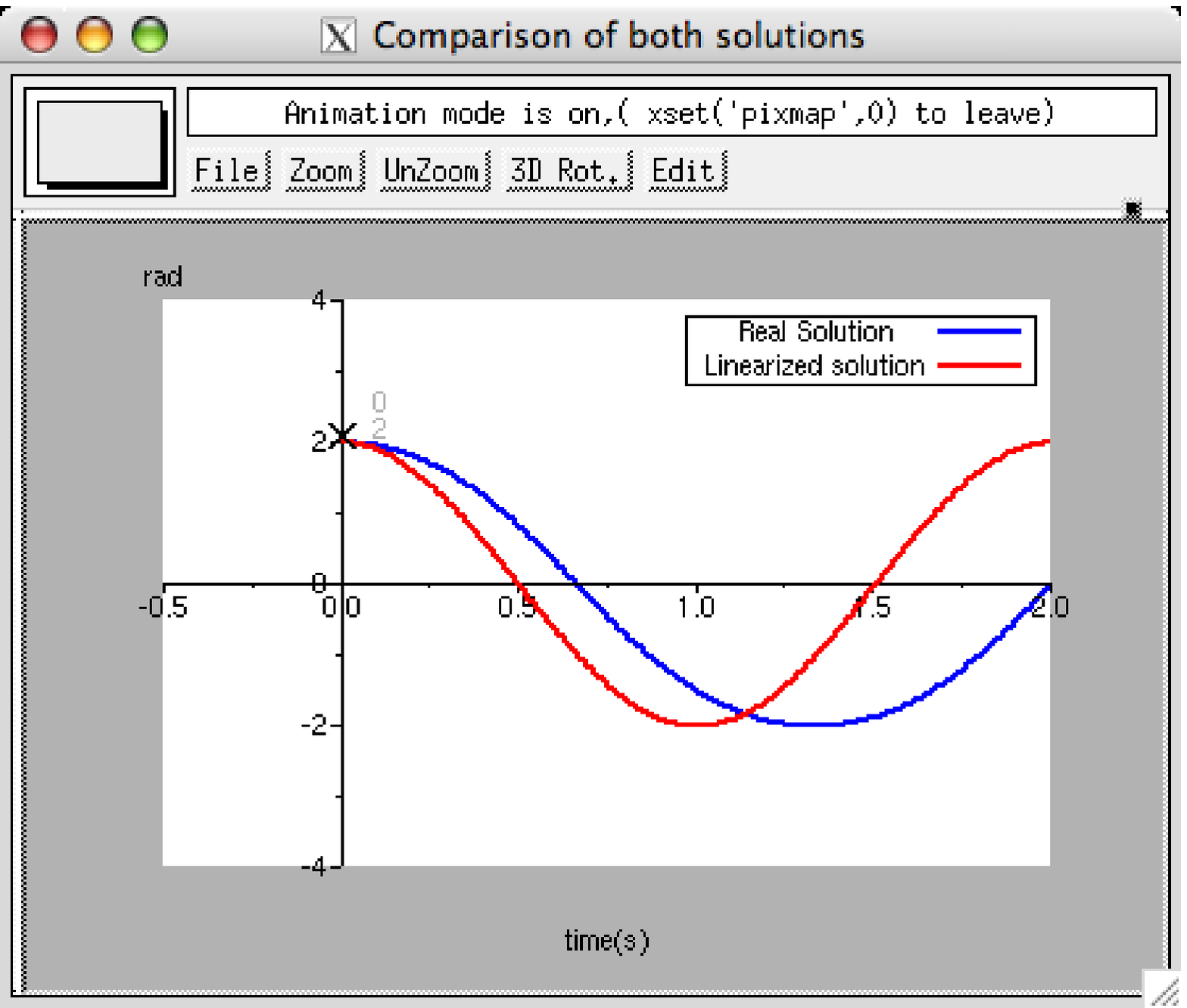}\end{minipage}\\
\end{array}
$$
\caption{Screenshots of the various tabs of the pendulum simulation running on a local client with Scilab. In the graphical window, the user can move
the cross shaped pointer and see the trajectory changes in real time.}
\label{scilab}
\end{figure}
\subsection{Offline batch publication of HTML pages}
\begin{figure}
$$
\includegraphics[width=0.75\columnwidth]{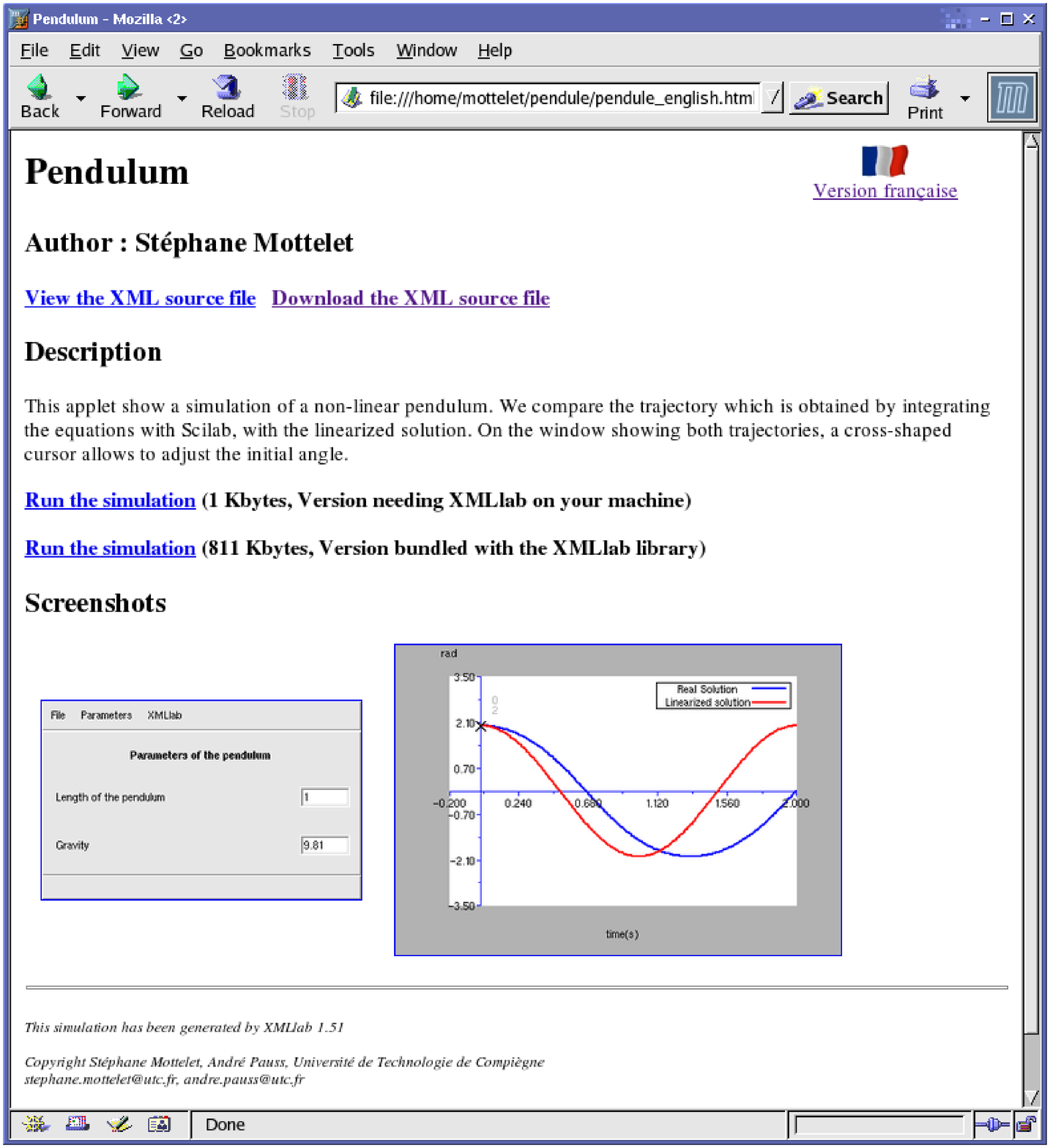}
$$
\caption{An HTML page describing the pendulum simulation. This kind of page is dedicated
to serve the simulations to clients where Scilab is installed}
\label{publish}
\end{figure}

When a large number of simulations have to be deployed in an educational context, it is possible
to automatically generate a tree of HTML files presenting the simulations e.g. sorted by categories. The HTML
files, as well as some screenshots
of the graphical output and of the user interface are automatically generated in a batch process which only takes
a few minutes. The user can then browse the different pages, take a look at the screenshots, read the description
and finally launch the chosen simulation. The typical HTML page for a single simulation is given on figure \ref{publish}. 
Note: this kind of publication still needs Scilab on the client machines (see the examples section of the XMLlab WWW site \cite{xmllab}).

\subsection{Online distant publication using the XMLlab WebServer}
Still in an educational context, but when it is not possible to have Scilab installed on all the client
machines, it is possible to publish the simulation towards thin clients running a simple WWW browser.
The drawbacks of such an approach are already known: there is an interactivity loss, but simulations can
be deployed very fast. 

\begin{figure}
$$
\includegraphics[width=0.6\columnwidth]{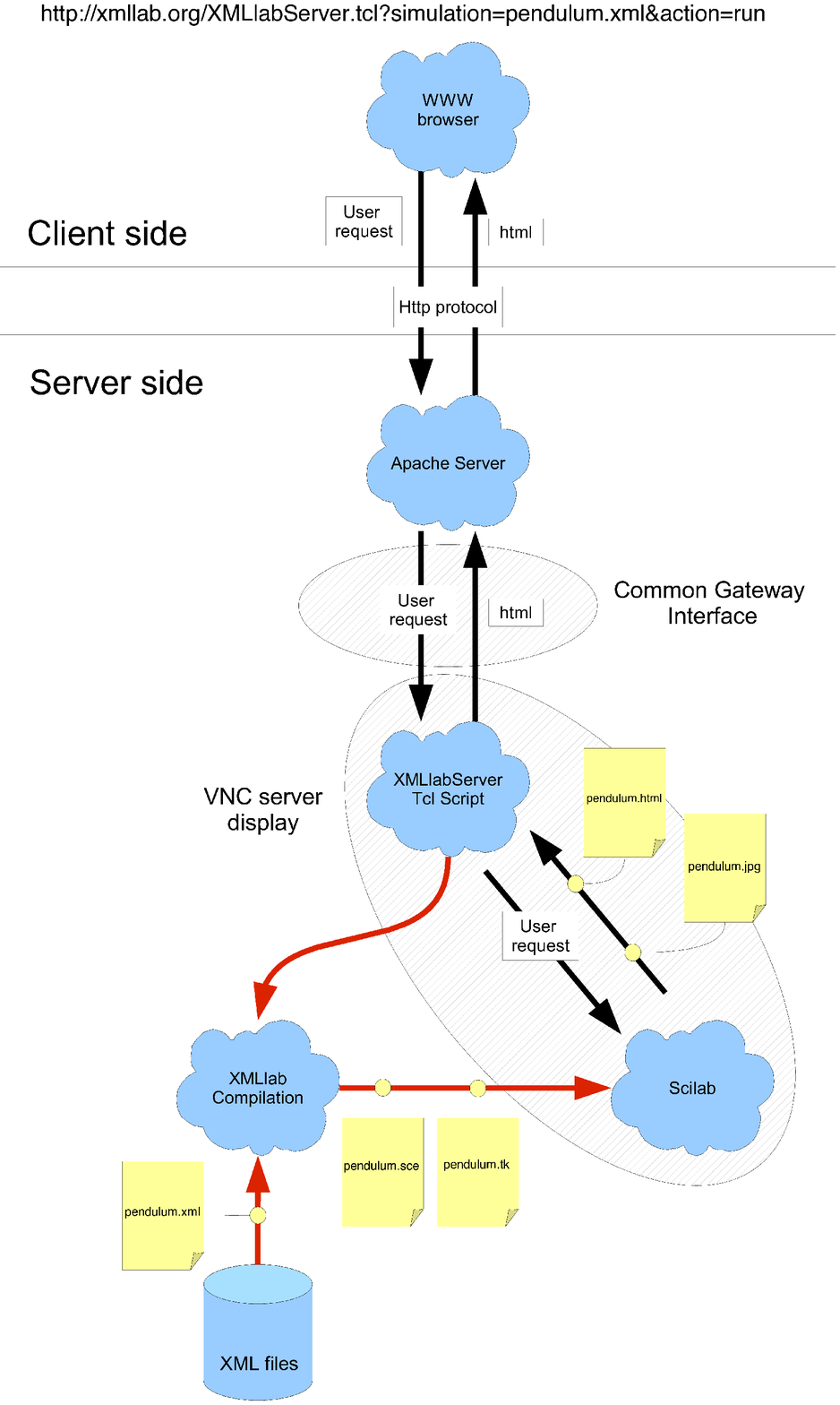}
$$
\caption{Internals of the XMLlab WebServer. Red arrows occur only at first user request.}
\label{server}
\end{figure}

We have chosen a rather classical architecture based on a HTTP server and the Common Gateway Interface. 
The entry point is a CGI script (written in Tcl) which processes the user requests (see figure \ref{server}). The
collection of XML simulations to be served are stored in the server (running any flavor of Unix) and the only needed software is:
\medskip\begin{itemize}
\item Scilab with the XMLlab toolbox,
\item An HTTP server, e.g. the Apache HTTP server,
\item The VNC virtual X11 server,
\end{itemize}
\begin{figure}
$$
\includegraphics[width=0.75\columnwidth]{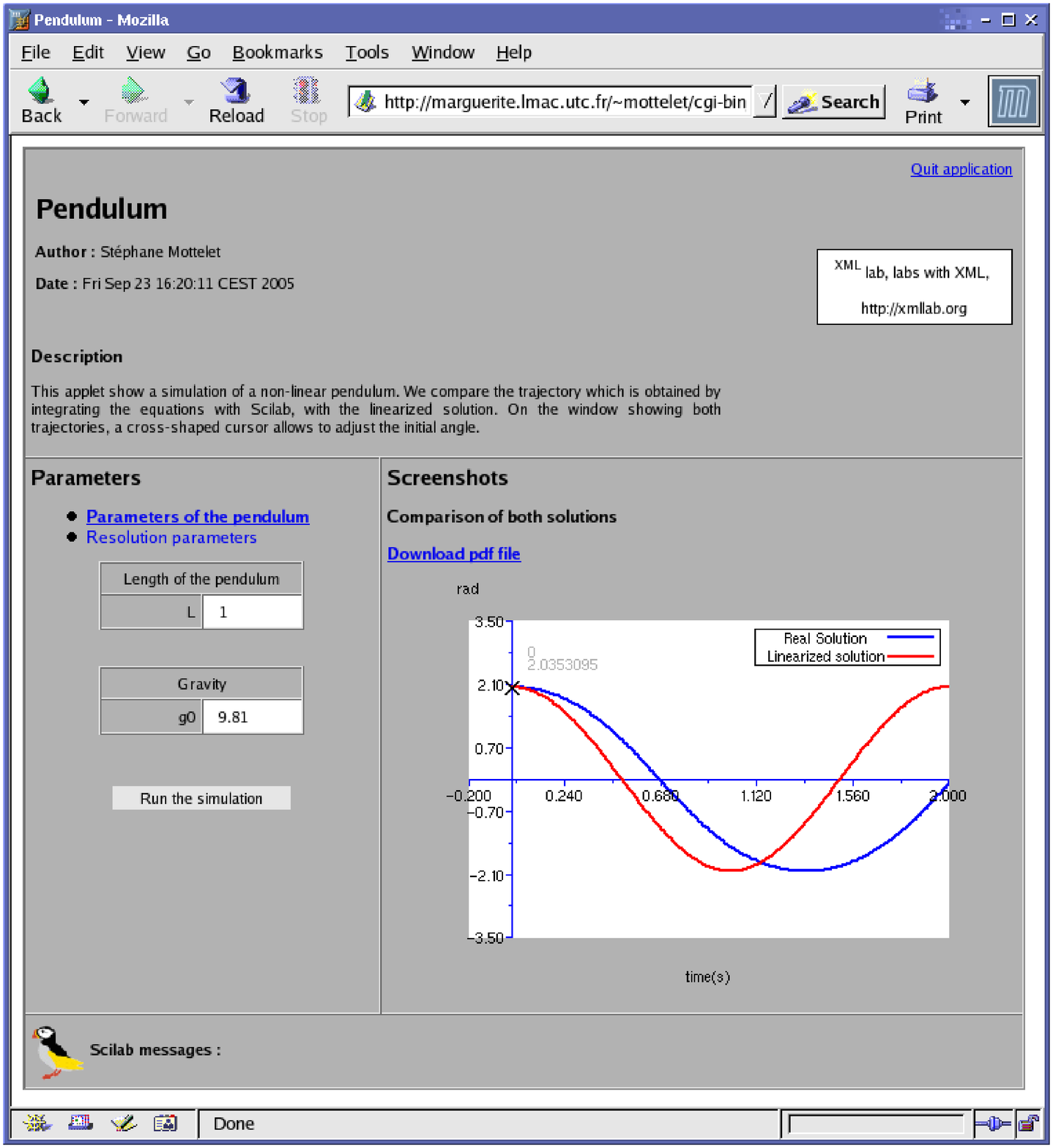}
$$
\caption{Screenshot of the pendulum simulation served by the XMLlab WebServer on a client machine running the 
Mozilla web browser on a Linux machine.}
\label{web}
\end{figure}
For the pendulum example, the generated HTML page can be seen on figure \ref{web}. The user can interact by changing the parameters values, 
browse the different parameter sections and download a PDF version of the graphical output.

\medskip We now describe how a typical URL (see on top of figure \ref{server}) is processed : when the first user request is processed, the CGI script  associates a session number to the client and performs the following tasks :
\begin{enumerate}
\item Retrieve the XML file (here \verb1pendulum.xml1) and process it in the XMLlab compilation chain. This produces
two files, a Scilab file \verb1pendulum.sce1 (dedicated to computations and graphical output) and a Tk file \verb1pendulum.tk1 (dedicated to
communications with the CGI script and HTML output).
\item Verify if the X11 VNC server is running (if not, a new server is launched), and launch a Scilab instance which uses this display
for its graphical output. 
\item Run  \verb1pendulum.sce1 and \verb1pendulum.tk1 into Scilab. The obtained result is an HTML file \verb1pendulum.html1 together with
image files corresponding with the first run of the simulation with default values of the parameters.
\item The HTML output is sent back to the client's WWW browser. 
\end{enumerate}

The red arrows in figure \ref{server} denote tasks which are only done at first user request. When the user interacts with the simulation,
then parameter changes are sent to the simulation running into Scilab by using the Tk send mechanism.

\section{Trends and conclusions}

The different parameters types and the mathematical objects presented in section 2 are already present in \xmllab\ 1.6,  
but \xmllab\ is a work in constant progress, and many extensions and improvements are necessary. However, we think
that the choices we have made are valid, especially concerning the structuring of the simulations and the 
architecture of the compilations chain. The needed development time to add a new type of equation is always limited: for 
example, the \verb1stationary-pde1 element, allowing to describe an elliptic partial differential 
equation (extension of the DTD and associated XSL stylesheet sections), has been developed in two days 
(we rely on a Scilab ``PDE toolbox'').

The XMLlab WebServer allowing dynamic publication of the simulation towards thin clients is the most recent
feature we have developed in \xmllab. Since we have opened the possibility to encapsulate Scilab scripts 
(making only computations) the XMLlab WebServer feature is giving to Scilab the equivalent 
of what the Matlab WebServer (this product is discontinued) used to provide to Matlab, but
with a completely different approach, since the user doesn't have to write any line of HTML. In this case,
complex Scilab scripts, eventually computing some new values of the parameters (see e.g. the linear regression example
in the Mathematics section of \xmllab\  examples) are embedded in a \verb1script1 element, replacing the \verb1compute1
element. Of course all the other high level elements are present, and allow to do the same multi-medium
publication as for pure XML simulation files (see e.g. the Discrete Cosine Transform and the Commutation Angles simulations
in the examples).

The future developments of \xmllab\ mainly focus on, discrete time systems simulation, stochastic systems, generation
of Adobe Flash animations, sound output, and so on. As far as the edition of the XML files is concerned, we plan to use ``Cascading Stylesheets'' 
allowing to edit XML files in a very user-friendly way in the XXE editor (see \cite{css2}). We also plan to migrate the 
actual DTD to an XML Schema (\cite{schema}), to allow some enhancements in the control of validity of the different data 
types contained in elements and attributes. In the current \xmllab\ release, this control is made in the XSL stylesheets.

At the time we are writing this paper,  \xmllab\ has already been used for 3 years in chemistry courses at the UTC (150 students by semester), 
under the form of demonstrations during the course, and during the labs, together with experimental acid/alkali titrations. With 
the help of simulation the students interpret the experimental curves and are able to answer reasoning questions. The software is
also available in the UTC intranet to allow students to improve their understanding of phenomenons. A sample survey has been made by 
e-mail at the end of each semester and allowed us to conclude that the software is easy to use, and students do not encounter major 
technical problems. The software allows a better understanding of complex phenomenon and acid/alkali equilibrium, but high level labs 
assistants are needed (they must be trained on the software before the labs). \xmllab\ is also used at the University Of Picardie Jules 
Verne since 2007, in the systems biology courses.

\xmllab\ has also been integrated in the prize-winning SCENARI Platform editorial chain (SCENARI Platform is an open source application suite
 for designing digital editing chains, used for creating professional standard multimedia documents, see \cite{scenari,scenariwww}). 
 
Of course, many other examples of simulations have been written to show that \xmllab\ can be used in various disciplinary fields :
we have examples in the fields of biology (microbial and enzymatic kinetics), physics (pendulum, oscillators, two-body 
system, Poisson equation, etc.), chemistry (Acid/Alkali equilibriums, chemical kinetics), chemical engineering (ideal reactors), etc. 
The complete list of available examples is given in appendix. All these simulations are available in French, English and Spanish, and 
the translation to German is in progress. We hope that a lot of people will contribute and request some 
new features, which will help us to make \xmllab\ fit to new disciplinary fields.

\xmllab\ is available at the address \verb1http://xmllab.org1, as 
 a Scilab toolbox under the GPL license. The distribution is available for
all popular platforms (Windows, Linux, Solaris, Mac OS X) and well integrated, e.g. 
the applets can be run directly by double clicking the icon of the file, without
having to launch Scilab before.

The documentation is available in French and English, under the form of a
quick start guide and a reference manual. 

\medskip \xmllab\ has been accepted as a contribution by the Scilab Consortium
in June 2004, and one of the two authors is now an official contributor member of the Consortium, also
sitting at the steering committee.

\section*{Acknowledgments}
This work has been financed by the groupment ``Evaluation of New 
Technologies in Education'' of the regional Council of Research from 
Picardie and by the UNIT consortium (French Numeric University of Engineering and Technology). IUFM's teachers and Jules Vernes University of Picardie 
Professors have contributed efficiently to the trandisciplinarity of 
this project.

\appendix 

\section{List of the simulations examples available at the XMLlab WWW site: xmllab.org}
\begin{itemize}
 \item Image Processing
\begin{itemize}
\item Discrete Cosine Transform.
\end{itemize}
\item Engineering
\begin{itemize}
\item Commutation angles.
\end{itemize}
\item Physics
\begin{itemize}
\item Damped Oscillator. Pendulum. Pendulum (with animation). Earth-Moon system. Simulation of the Laplace equation.
\end{itemize}
\item Maths
\begin{itemize}
\item Lissajous curve. A simple surface example. Tangent. Osculating polynomials. Gaussian mix. Lagrange Interpolation. Cycloïd. Linear regression.
Inversion of a matrix. Helix. System of differential equations.
\end{itemize}
\item Predation
\begin{itemize}
\item Kinetics of prey predation by predators. Kinetics of rabbit predation by foxes.
\end{itemize}
\item Chemical kinetics
\begin{itemize}
\item Nonreversible reaction. Reversible reaction(equilibrium). Simultaneous reactions (in parallel). Successive reactions with an equilibrium.
\end{itemize}
\item Enzymatic kinetics.
\begin{itemize}
\item Enzymatic kinetics with a competitive inhibition. Enzymatic kinetics with a incompetitive inhibition. Enzymatic kinetics with a non-competitive
inhibition. Michaelis-Menten's enzymatic kinetics.
\end{itemize}
\item Microbial kinetics
\begin{itemize}
\item Graef-Andrews's growth kinetics. Monod's growth kinetics.
\end{itemize}
\item pH titrations
\begin{itemize}
\item Simulation of acid-alkali titration in water. Simulation of a single acid titration.
\end{itemize}
\end{itemize}

\end{document}